\newcommand\bm[1]{\mbox{\boldmath $#1$}}
\newcommand\Eq[1]{Eq.$\:$[\,\ref{#1}\,]}
\newcommand\Fig[1]{Fig.$\:$\ref{#1}}
\newcommand\Figure[1]{Figure$\:$\ref{#1}}
\newcommand\Equation[1]{Equation$\:$[\,\ref{#1}\,]}

\newcommand\w[1]{\mbox{$\omega_{#1}$}}
\newcommand\boldw[1]{\mbox{$\bm{\omega}_{#1}$}}

\newcommand\dm{$\rho$}

\documentclass[aps,pra,reprint,groupedaddress,showpacs,amsmath,amssymb,floatfix]{revtex4-1}

\usepackage{graphicx}
\usepackage{bm}

\begin{document}



\title{Further perspective on the theory of heteronuclear decoupling}

\author{Thomas E.~Skinner}
\email{thomas.skinner@wright.edu}  
\affiliation{Physics Department, Wright State University, Dayton, OH 45435}

\date{\today}

\begin{abstract}
An exact general theory of heteronuclear decoupling is presented for spin-1/2 IS systems. RF irradiation applied to the I spins decouples the S spins by reversing the sign of antiphase magnetization.  A one-to-one correspondence is demonstrated between the sign of $2S_yI_z$ and the sense of the S-spin coupling evolution, with magnetization $S_x$ refocused when $\int 2S_yI_z\,dt = 0$.  The exact instantaneous coupling at any time during the decoupling sequence is readily obtained in terms of the system states, showing that the creation of two-spin coherence is crucial for reducing the coupling, as required during the refocusing process.  Representative examples from standard cyclic, periodic composite-pulse and adiabatic decoupling sequences illustrate the decoupling mechanism.  More general aperiodic sequences, obtained using optimal control, realize the potential inherent in the theory for significantly improved decoupling.  The recently derived equivalence between the dynamics of any N-level quantum system and a system of classical coupled harmonic oscillators provides additional insight into the mechanics of decoupling.
\end{abstract}



\maketitle

\section{Introduction}
The dominant strategy for designing broadband heteronuclear decoupling sequences for the past 30 years derives from average Hamiltonian theory \cite{AvgHamTheory1,AvgHamTheory2}.  If RF irradiation is applied to the I spins in a weakly coupled spin-$\frac{1}{2}$ IS system, effective decoupling of the S spins can be shown to occur if the I spins are cyclically returned to their initial state.  The range of chemical shift offsets over which this is possible defines the bandwidth of the decoupling sequence.

At the same time, Waugh's theory of decoupling \cite{Waugh82} showed that the cyclic condition is not crucial.  The magnitude of the net rotation that a free (uncoupled) I-spin would undergo during a decoupling period only has to vary sufficiently slowly over the bandwidth.  
The variation of this net rotation with offset is small enough if it is approximately constant over frequency ranges equal to the scalar coupling.
A net rotation of zero over the entire bandwidth, as required for cyclicity, is unnecessarily restrictive.  

However, as emphasized in \cite{Waugh82}, the theory provides no insight or procedure for designing effective noncyclic sequences.  In addition, the theory has its own restrictions.  Periodic irradiating sequences are assumed in the analysis, with one sampling point per period.   Ideal decoupling  is obtained if the in-phase magnetization at each sample is equal to its initial value (nominally, one).  Deviations from this ideal represent modulations of the signal that appear as satellite lines or sidebands.  Smaller deviations result in smaller sidebands. 
Sampling faster than the assumed rate, as in practical decoupling applications, was not considered critical, since it merely makes the sidebands more prominent.  These restrictive assumptions were therefore justified on qualitative grounds, while recognizing their potential limitations with regard to effective decoupling.  A more general theory for deriving good decoupling schemes was posed as a challenge for further consideration.

Thus, for lack of a better alternative, 
decoupling strategies have emphasized cyclic sequences.  
The standard approach is to find the best inversion pulse based on a set of desired criteria (power, bandwidth, etc.) and apply phase cycles to approach an ideally cyclic sequence \cite{Grutzner1975173, Basus197919, LFF82, LEVITT198347, Shaka83a, Shaka1985547, Shaka198747, Geen19961699, Starcuk199424, Bendall1995126, Kupce1996299, Fu1996129, Levitt1982157, Tycko1984462, Tycko19852775, Tycko19861905, Fujiwara198853, Fujiwara1990584, Fujiwara1993103, PhCycAlgAdiDec97}.  A noteworthy example is the adoption of adiabatic pulses for the inversion element of the decoupling sequence \cite{Basus197919, Starcuk199424, Bendall1995126, Kupce1996299, Fu1996129}, resulting in broadband inversion at lower power than composite pulses.  There are a multitude of adiabatic pulses that can provide the necessary bandwith for high-field spectroscopy at a given average pulse power level. They differ in the intensity of sidebands produced over the decoupled bandwidth \cite{PeakPwrEff96}. Performance limits for ideal adiabatic decoupling have been derived that provide a standard for decoupling performance \cite{STUD+Calib98}, but are themselves limited by the framework of cyclic decoupling strategies.

In the present work, a unifying mechanism for decoupling is derived, with no assumptions about the sequence itself.  An exact, time-dependent coupling emerges as a logical consequence.  These results clarify possibilities for improved decoupling, unconstrained by the demands of periodic and cyclic sequences.  Optimal control is able to realize this potential and, in effect, invert the theory to derive more effective decoupling schemes \cite{Neves09, BUSS}.  The fidelity of the derived decoupling mechanism is illustrated using both standard cyclic sequences and this new generation of improved, aperiodic decoupling sequences.

\section{Theory}
A brief summary of the existing theory of heteronuclear decoupling is provided first to establish the context of standard cyclic decoupling strategies. This is followed by a new perspective on decoupling that augments Waugh's original treatment and provides the mechanism underlying decoupled IS systems.

\subsection{The decoupled signal}
\label{subsec:DecSignal}
The Liouville-von Neumann equation for the time evolution of a density matrix \dm\ governed by system Hamiltonian $\cal H$ is 
     \begin{equation}
\dot\rho  = -i\,[\,{\cal H},\rho\,].
\label{Liouville}
     \end{equation}
For time-independent Hamiltonian, the formal solution  is 
     \begin{eqnarray}
\rho(t)  &=& e^{-i\,H t}\,\rho(0)\, e^{i\,H t} \nonumber \\
         &=& U \rho(0)\, U^\dag,
\label{SolLiouville}
     \end{eqnarray}
which defines the propagator $U(t) = e^{-i\,H\, t}$.  

Constant RF irradiation of amplitude $\omega_{\mathrm{rf}}$ and phase $\phi$, or, $(x,y)$ components $(\omega_1, \omega_2)$, and offset $\omega_3$ relative to the resonance frequency of the I spins results in an effective field in the rotating frame of the I spins written (in angular frequency units) as
     \begin{eqnarray}
\boldw{e} &=& \w{\mathrm{rf}}[\,\cos\phi\,\hat{\bm{x}} \,+\, 
               \sin\phi\,\hat{\bm{y}}\,] \,+\, \omega_3\,\hat{\bm{z}} 
          \nonumber \\
          &=& \omega_1\, \hat{\bm{x}} + \omega_2\, \hat{\bm{y}} +
              \omega_3\, \hat{\bm{z}}.
\label{w_eff}
     \end{eqnarray}
In the weak coupling limit, the Hamiltonian for on-resonance S spins is
     \begin{equation}
{\cal H} = \boldw{e} \cdot \bm{I} + \mathcal{J} S_zI_z,
\label{Ham}
     \end{equation}
where $\cal J$ is $2\pi$ times the coupling $J$ in Hertz.  $\cal H$ can be described in terms of equivalent effective fields
$\boldw{\pm} = \boldw{e} \pm {\cal J}/2\, \hat{\bm{z}}$, giving a propagator comprised of
$U_\pm(t) = e^{-i\, (\boldw{\pm}\cdot\bm{I}) t}$ \cite{Waugh82}.  

Since $[S_z,{\cal H}]=0$, the offset effect of spin S can be calculated separately. It is therefore sufficient to consider the S spins on resonance, with solution \cite{Waugh82}
     \begin{eqnarray}
\hskip -0.1truein
S_x(t) &=& \frac{(1 + \hat{\bm{\omega}}_+ \cdot \hat{\bm{\omega}}_-)}{2}\,
                    \cos\frac{(\w{+}t - \w{-}t)}{2}   \nonumber \\
       &+& \frac{(1 - \hat{\bm{\omega}}_+ \cdot \hat{\bm{\omega}}_-)}{2}\,
                    \cos\frac{(\w{+}t + \w{-}t)}{2} ,
\label{S_x(t)}
     \end{eqnarray}
where $\hat{\bm{\omega}}_\pm$ are unit vectors in the direction of the equivalent effective fields.

\subsection{Standard decoupling strategy: an I spin emphasis}
\Equation{S_x(t)} forms the basis for an interpretation of decoupling in terms of I spin evolution \cite{Waugh82}, summarized here as follows.  A shaped RF pulse is typically applied in discrete steps of length $\Delta t$ at succesive times $t_k = k\Delta t$ during which the amplitude $\w{\mathrm{rf}}^{(k)}$ and phase $\phi^{(k)}$ are constant.
Each propagator $U_\pm^{(k)}$, if operating on an uncoupled I spin during interval $k$, would generate a rotation about respective axes $\hat{\bm{\omega}}_\pm^{(k)}$ by angles 
$\beta_\pm^{(k)} = \omega_\pm^{(k)}\Delta t$.  At time $t_n$, the $U_\pm(t_n)$ are 
the concatenation of the successive propagators $U_\pm^{(k)}$ ($k = 1,\ldots,n$). The solution for $S_x(t_n)$ is then of the form given in \Eq{S_x(t)}, with
$\w{\pm}t_n = \beta_\pm$ denoting equivalent net rotation angles about axes 
$\hat{\bm{\omega}}_\pm$, which can be determined by expanding the exponential in $U_\pm(t_n)$ to obtain
     \begin{equation}
U_\pm(t_n) = \cos(\beta_\pm /2) - i\,(\bm{\hat\omega}_\pm \cdot \bm{I}) \sin(\beta_\pm /2).
\label{U_pm}
     \end{equation}

As noted in \cite{Waugh82}, if the sequence is cyclic at time $t_n$ in the sense that $\beta_\pm \approx 0$, then 
$\hat{\bm{\omega}}_+ \approx \hat{\bm{\omega}}_-$ and $S_x(t_n) = 1$.  There is therefore no net coupling evolution, and S appears to be decoupled from the I spins.  The second term in \Eq{S_x(t)} is small under these conditions, responsible for the production of sidebands, and was generally ignored.
Since the offset for each $\omega_\pm^{(k)}$ is equal to 
$\omega_3 \pm {\cal J}/2$, an equivalent interpretation is that the propagator for a free, uncoupled I spin (${\cal J} = 0$ in \Eq{Ham}) should return the I spins to their initial state, independent of offset.  Then $\beta_\pm$ is automatically zero over the given range of offsets, which defines the bandwidth for good decoupling.  

This is the standard approach to developing good broadband decoupling sequences, proposed, prior to Waugh's analysis, by Levitt and Freeman \cite{Levitt81} using average Hamiltonian theory.  Other options for achieving $S_x(t_n)=1$ do not lend themselves to such an intuitive picture.  One possibility emphasized in \cite{Waugh82} for periodic, but not necessarily cyclic, sequences is $\beta_+ \approx \beta_-$ for rotation axes $\hat{\bm{\omega}}_+ \approx \hat{\bm{\omega}}_-$.  This is equivalent to requiring that the net rotation of a free I spin during a given $t_n$ be insensitive to offset.  Although this is more general than the cyclic criterion of net zero rotation, it is less clear how to achieve it.

There are other possibilities, as well, that justify sampling at times 
$t \leq t_n$ and don't require periodicity in the sequence. If $\hat{\bm{\omega}}_+$ and  $\hat{\bm{\omega}}_-$ are nearly perpendicular, then for sufficiently small (but not necessarily equal) $\beta_\pm = \omega_\pm t$, the cosine terms in \Eq{S_x(t)} are each approximately equal to one, giving 
$S_x(t)\approx 1$.  Solutions satisfying this set of conditions (see, for example, \Fig{DecParams}) can be found, ironically, in decoupling sequences designed according to the cyclic standard $\beta_\pm = 0$, putting an exclamation point on Waugh's insight that good decoupling does not require cyclicity.  In addition, they also highlight the  significance of the second term in \Eq{S_x(t)}, which most generally is not small relative to the first term.

There are, in principle, innumerable combinations of $\hat{\bm{\omega}}_\pm$ and $\beta_\pm$ that give a solution $S_x(t) \approx 1$.  Further detail is provided in $\S$\ref{sec:Examples}.  A more general approach to decoupling than the either the cyclic or periodic standard might determine some number of these combinations and set about devising I spin propagators to access them. If this is not difficult enough, one should also be able to distinguish between the relative merits of various combinations (most of which remain unknown) as the propagator is being derived.  One immediately understands and appreciates the exclusive focus on cyclic decoupling methods during the 30 years since Waugh suggested there should be other possibilities. 

Shifting focus to the S spins and evolution of the states given by the complete solution for $\rho(t)$ \cite{Skinner99,Bendall00,Skinner02} provides a more fruitful approach for understanding and developing improved decoupling sequences.

\subsection{New perpectives on decoupling: an S spin emphasis}
Initial $\rho(0) = S_x$ also evolves to states $2S_yI_j$, 
($j = 1,2,3 \equiv x,y,z$) which, although not directly detectable, are important for understanding the mechanism of decoupling.  The Hamiltonian of \Eq{Ham}, in fact, restricts the time evolution of operators in the set $\{S_x, 2S_yI_j\}$ to the subspace spanned by these same opertors, and similarly for the set $\{S_y, 2S_xI_j\}$ \cite{Skinner99}, as can be seen by calculating the commutator in \Eq{Liouville}.  Complete solutions for the evolution of any state in either closed set to the other states in the respective set are given in \cite{Skinner99,Bendall00,Skinner02}.  Defining $S_+ = S_x + i\,S_y$ and writing the identity element as $I_0$ reduces the solution for $\rho(t)$ in \Eq{SolLiouville} to a calculation of
     \begin{equation}
\rho \sim S_+ (U_+ I_\nu U_-^\dag), \qquad \nu = 0,1,2,3.  
\label{ExactEvol}
     \end{equation}
The evolution of the system states is distinctly \textit{not} a rotation of the I spins, which would be of the form $U I U^\dag$.  

As noted, an analysis of decoupling in terms of the equivalent rotation of free, uncoupled I spins does give the correct answer at the specific times when the sequence is cyclic.  In addition, \Eq{ExactEvol} shows that this analysis will also be correct if $U_+ = U_-$, which becomes a reasonably good approximation at sufficiently high power.  A lower boundary 
$\omega_\mathrm{rf} \approx 50{\cal J}$ for the high-power approximation has been found to limit simulation or spectral errors to less than 5\% \cite{Bendall00}.  For $J = 150$ Hz, this requires an RF field $\gtrsim 7.5$ kHz.  However, significantly lower power levels are typically used for decoupling, due to sample heating constraints.  

Thus, the full, exact solution is required for a comprehensive analysis of decoupling.  It provides considerable latitude for improving decoupling performance, freed from the too restrictive constraint of making the sequence cyclic over large bandwidths. Requiring periodic decoupling sequences that refocus in-phase magnetization at the end of each period is also unnecessary and can be abandoned.

\subsubsection{The decoupling mechanism}
\label{subsubsec:DecMech}
Expressing \dm\ as a linear combination of the basis states $\{S_x, 2S_yI_j\}$, plugging into \Eq{Liouville}, and equating like terms gives
the equation of motion for the coefficients $\bm{\tilde{r}} = (r_0,r_j)$, which are the expectation values of the associated operators
$\{S_x, 2S_yI_j\}$.  Alternatively, one can successively assign each member of the basis set to $\rho$ in \Eq{Liouville} and reverse the sign of the result to reflect the equivalence between evolving the basis states or evolving the components in the opposite sense.  Either way, one obtains 
\begin{subequations}
\label{rdotEq}
     \begin{eqnarray} 
\hskip -18pt \frac{d}{dt}\,{\bm{\tilde{r}}}(t) &=& \Omega\, \bm{\tilde{r}}(t) \label{rdotEq:1} \\
\ &\ & \nonumber \\
\Omega &=& \left( \begin{array}{cccc}
                       0 & 0 & 0 & -\w{_J} \\
                       0 & 0 & -\omega_3 & \omega_2 \\
                       0 & \omega_3 & 0 & -\omega_1 \\
                      \w{_J} & -\omega_2 & \omega_1 & 0
                    \end{array}
             \right) ,  \label{rdotEq:2}
     \end{eqnarray}
\end{subequations}
with $\w{_J} \equiv \mathcal{J}/2$ for a compact rendering of $\Omega$.
The solution $\bm{\tilde{r}}(t) = e^{\Omega t} \bm{\tilde{r}}(0)$ has been calculated using a different formulation of the problem and presented in several formats \cite{Skinner99,Bendall00,Skinner02} which give, by inference, the matrix exponential.

The S magnetization can be represented in the usual fashion by two equal components $S_\alpha$ and $S_\beta$ aligned initially along the $x$ axis.  They are coupled, respectively, to I spins that are either up or down along the $z$ axis and precess in opposite directions on resonance. \Equation{rdotEq} shows immediately that $r_0$ changes at a rate $-\w{_J} r_3(t)$, giving (more explicitly)
      \begin{equation}
\dot S_x(t) = -\left(\frac{\mathcal{J}}{2}\right) 2 S_yI_z(t).
\label{SxdotEq}
      \end{equation}

Thus, the \textit{fundamental decoupling mechanism} is the change in sign of antiphase magnetization to repeatedly refocus $S_x$, which increases or decreases as $2S_yI_z$ is negative or positive, respectively. 

Integrating \Eq{SxdotEq} between times $t_1$ and $t_2$ when $S_x \approx 1$ gives $S_x(t_2) - S_x(t_1) \approx 0$, so the average value of 
$2S_yI_z \approx 0$ over the same time span.  Thus, starting at $t = 0$ when $S_x = 1$, the most efficient decoupling will subsequently average $2S_yI_z$ to zero over the shortest possible time spans for a given power level, thereby maximizing the average value of $S_x$ over the same time intervals.  

This is an exact and more precise criterion for good decoupling compared to average Hamiltonian methods, which are designed to average the toggling frame coupling term to zero.  It is also more general than the criterion defined in terms of an effective coupling parameter in \cite{Waugh82}, as shown in section $\S$\ref{subsubsec:Waugh_lambda}.

\subsubsection{Equivalent harmonic oscillator model}

The dynamics of any N-level quantum system can be represented by a system of classical coupled harmonic oscillators \cite{QM-SHO}.  In particular, this equivalence is applicable to the spin systems in magnetic resonance and provides additional insight on decoupling.  Differentiating \Eq{rdotEq:1} gives
\begin{subequations}
\label{rddotEq}
     \begin{eqnarray} 
\frac{d^2}{dt^2}\, {\bm{\tilde{r}}}(t) & = &
      \Omega^2\, \bm{\tilde{r}}(t) \label{rddotEq:1} \\
\ &\ & \nonumber \\
\Omega^2 & = &   \label{rddotEq:2} \\
 & & \hskip -16pt
 \left( \begin{array}{cccc}
 -\omega_{_J}^2 & \w{_J}\w{2}  & -\,\w{_J}\w{1} & 0 \\
 \w{_J}\w{2} & -\,\omega_{23}^2 & \w{1}\,\w{2} & \w{1}\,\w{3} \\
 -\,\w{_J}\w{1} & \w{1}\,\w{2} & -\,\omega_{13}^2 & \w{2}\,\w{3} \\
 0 & \w{1}\,\w{3} & \w{2}\,\w{3} & -\,\omega_{12,J}^2
                    \end{array}
             \right) , \nonumber
     \end{eqnarray}
\end{subequations}
with $\omega_{ij}^2 \equiv \omega_i^2 + \omega_j^2$ and 
$\omega_{12,J}^2 \equiv \omega_{12}^2 + \omega_{_J}^2$ (again, to keep the matrix relatively compact).  \Equation{rddotEq} describes four unit masses coupled by springs of stiffness $k_{ij} = (\Omega^2)_{ij}$ and natural frequencies determined by the self-couplings 
$k_{ii} = -(\Omega^2)_{ii} - \sum_{\,l\, \neq\, i} |(\Omega^2)_{il}|$.
Associating a two-component vector $\bm{\omega_{ij}} \equiv (\w{i},\w{j})$ with $\omega_{ij}^2$ and defining $\bm{r_{ij}}$ similarly gives
\begin{subequations}
\label{r_muddotEq}
     \begin{eqnarray} 
\ddot{r}_0 + \omega_{_J}^2 \,r_0 & = & 
       -\,\w{_J}(\bm{\omega_{12}} \times \bm{r_{12}}) \label{r_muddotEq:1} \\
\ddot{r}_1 + \omega_{23}^2 \,r_1 & = & 
       \w{2}\,(\w{_J} r_0) + \w{1}(\bm{\omega_{23}}\cdot \bm{r_{23}}) \label{r_muddotEq:2} \\
\ddot{r}_2 + \omega_{13}^2 \,r_2 & = & 
       -\,\w{1}\,(\w{_J} r_0) + \w{2}(\bm{\omega_{13}}\cdot \bm{r_{13}}) \label{r_muddotEq:3} \\
\ddot{r}_3 + \omega_{12,J}^2 \,r_3 & = & 
       \w{3}\,(\bm{\omega_{12}} \cdot \bm{r_{12}}). \label{r_muddotEq:4}
     \end{eqnarray}
\end{subequations}

In the absence of applied RF, $\bm{\omega}_{12} = 0$, and one recovers the standard, \textit{J}-coupled sinusoidal evolution of $r_0 = S_x$ and $r_3 = 2S_yI_z$ at frequency $\omega_{_J} = \mathcal{J}/2$.  RF applied on-resonance still gives simple harmonic oscillation of $r_3$, but at the higher frequency
$\omega_{12,J} > \w{_J}$, which increases the refocusing rate of $r_0$ and thus decouples $S_x$ with increasing effectiveness as the RF amplitude increases.  In the on-resonance case, the components $(r_0, r_1, r_2)$ depend only on $r_3$, according to \Eq{rdotEq}. They are readily obtained for arbitrary initial condition $\bm{\tilde{r}}(0)$ by integrating the simple sinusoidal solution of \Eq{r_muddotEq:4} satisfying the associated initial condition $(d/dt)\bm{\tilde{r}}(0)$ obtained from \Eq{rdotEq}.  The on resonance solutions can also be found in \cite{Skinner99,Bendall00,Skinner02}.

More generally, \Eq{r_muddotEq} describes driven harmonic oscillators, with the driving force for $r_0$ and $r_3$ provided by two-spin coherence $\bm{r}_{12}(t)$.  Conversely, two-spin coherence in Eqs.~[\ref{r_muddotEq:2}] and [\ref{r_muddotEq:3}] is driven by in-phase and anti-phase magnetization.  
This mutual interaction among the operators, coupled by the RF irradiation, is the broader mechanism by which decoupling occurs.
The timely creation of two-spin coherence plays an important role in efficient decoupling, since, otherwise, $r_0 = S_x$ evolves at the natural coupling frequency $\omega_{_J}$ when $\bm{r}_{12} = 0$.  An exact expression for the modified coupling resulting from IS coherence is derived next.

\subsubsection{The instantaneous, time-dependent coupling}

Refocusing $S_x$ requires the apparent coupling to become zero and then change sign to reverse the precession of the associated in-phase components 
$S_{\alpha,\beta}$.  The exact (reduced) coupling frequency, $\mathcal{J}_r(t)/2 \leq \mathcal{J}/2$, at any time during a decoupling sequence is readily obtained in terms of the system states.

Defining $\varphi$ as the angle either component $S_{\alpha,\beta}$ makes with the $x$ axis gives $S_x = \cos\varphi$, so that 
$\dot S_x = -\dot \varphi \sin\varphi$.  But $\dot\varphi$ is the angular frequency $\mathcal{J}_r / 2$ at which the $S_{\alpha,\beta}$ precess about the $z$ axis, and $\sin\varphi = \sqrt{1-S_x^2}$.  Equating \Eq{SxdotEq} and the above expression for $\dot S_x$ gives an instantaneous reduced coupling
    \begin{eqnarray}  
{\cal J}_r(t) &=& {\cal J}\, \frac{2S_yI_z(t)}{\sqrt{1-S_x^2(t)}} \nonumber \\
              &=& {\cal J}\, \frac{2S_yI_z}
          {\sqrt{2S_yI_x^{\,2} + 2S_yI_y^{\,2} +2S_yI_z^{\,2}}}  
\label{J_r}
     \end{eqnarray}
for normalized states $||\bm{\tilde{r}}(t)|| = 1$, with 
$\mathcal{J}_r \rightarrow \mathcal{J}$ for $S_x \rightarrow 1$.

This expression quantifies the role of two-spin coherence in reducing the coupling and makes precise a geometrical vector description of coupling and system dynamics obtained previously \cite{Bendall00}.  It is readily integrated to obtain the average precession frequency $\bar{\mathcal{J}}_r/2$ during any interval $\Delta t_i = t - t_i$.  
Noting that $\dot{S}_x / \sqrt{1-S_x^2} = -\,(d/dt) \cos^{-1}S_x$, or, more intuitively, that $\bar{\mathcal{J}}_r/2 = \Delta\varphi / \Delta t$, with
$\varphi = \cos^{-1}S_x$, gives an average coupling 
     \begin{equation}
\bar{\mathcal{J}}_r(\Delta t_i) =  
   2\, \frac{\cos^{-1}S_x(t) - \cos^{-1}S_x(t_i)}
        {\Delta t_i}\,.
\label{AvgJ_r}
     \end{equation}

\subsubsection{A previous definition of the effective coupling}
\label{subsubsec:Waugh_lambda}
The scaling factor $\lambda$ defined for periodic decoupling sequences in \cite{Waugh82} is a special case of the current result.
During the decoupling period $\tau$ starting at $t_i = 0$, where 
$S_x = 1$, the effective coupling  is simply the average of the instantaneous coupling, which was scaled as
$\bar{\mathcal{J}}_r(\tau) = \lambda \mathcal{J}$.  A good approximation for $\lambda$ was derived for free I spins in terms of resonance offset $\delta$ and net rotation angle $\beta$ during the period $\tau$ as 
     \begin{equation}
\lambda\,(\tau) = (1/\tau)\,\partial\beta/\partial\delta ,
\label{Waugh_lambda}
     \end{equation}
which is exact in the limit $\mathcal{J} \rightarrow 0$.  More generally, 
\Eq{AvgJ_r} gives an exact expression for the scaling at any time during the decoupling period, valid for arbitrary decoupling sequences and any value of $\mathcal{J}$, as
     \begin{equation}
\lambda(t) =  \frac{\cos^{-1}S_x(t)}{({\cal J}/2)\,t}\,.
\label{lambda}
     \end{equation}
Although $\lambda$ might appear to get smaller (i.e., better decoupling) for \textit{larger} $\cal J$, $S_x$ also decreases nonlinearly with increasing $\cal J$.  
\begin{figure*}[t]
\includegraphics[scale=.95]{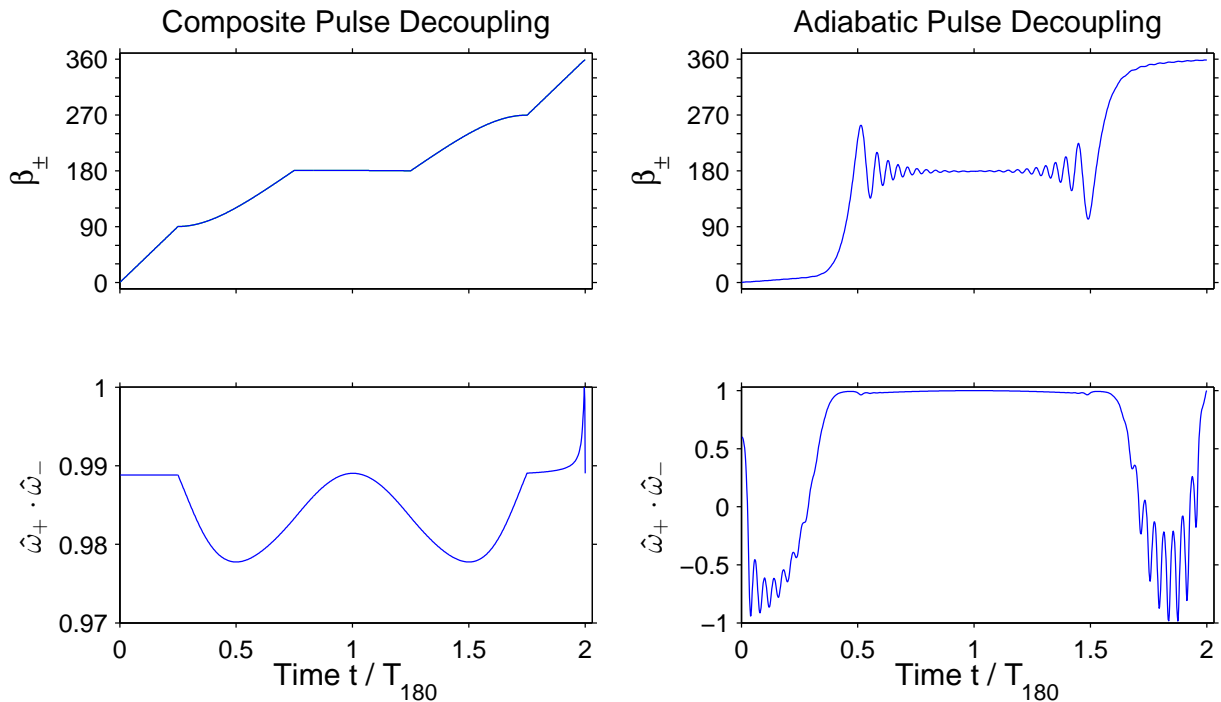}
\caption{Rotation parameters for a free (uncoupled) I-spin during irradiation by \textbf{(left):} two composite $90_x 180_y 90_x$ inversion pulses and \textbf{(right):}  two hyperbolic secant inversion pulses.  Rotation angles 
$\beta_\pm = \omega_\pm t$ for I-spins $\pm\mathcal{J}/2$ off resonance are plotted in the top panels as a function of time, relative to the inversion pulse length $T_{180}$ of each pulse.  RF amplitude $B_1 \gg \mathcal{J}/2$, so $\beta_+ \approx \beta_-$ during both pulses.  The dot product of the effective rotation axes $\bm{\hat{\omega}}_\pm$, plotted in the lower panels, and $\beta_\pm$ determine the decoupled signal $S_x(t)$ according to \Eq{S_x(t)}.
}
\label{DecParams}
\end{figure*}
In \cite{Waugh82}, the criterion for good decoupling is that $\lambda\,(\tau)$ be small for all offsets within the desired decoupling bandwidth. This requires the net propagator at each offset, the equivalent net rotation $\beta$, and $\partial\beta/\partial\delta$ at each offset, according to \Eq{Waugh_lambda}.
\Equation{lambda} only requires $S_x(\tau)$, which is equivalent to calculating the net propagator, at each offset. 

The utility of \Eq{Waugh_lambda} for evaluating decoupling sequences is considerably reduced in an era of faster computers.  The net propagator at the end of the decoupling period requires the net propagator at each preceeding time. The full FID can be quickly calculated at each offset, Fourier transformed, and evaluated to obtain the intensity of the centerband and sidebands---information not available using $\lambda$.  Nonetheless, an improved criterion for good decoupling would be that 
$\lambda(t)$, i.e., $\bar{\mathcal{J}}_r(t)$, be minimal at all times over the bandwidth, in addition to the end of the decoupling period $\tau$. This maximizes $S_x$ at each time over the bandwidth within the limits of what is possible, which is precisely the standard enforced in deriving improved decoupling sequences using the optimal tracking algorithm, considered next.

\subsection{Optimal control approach to decoupling}

Ideally, the theory would now be inverted to derive effective decoupling sequences.  Such inversions can be notoriously difficult.  Optimal control theory \cite{PinchOC} finesses the inverse solution by efficiently iterating trial forward solutions.  The initial state is evolved forward and compared at each time with the inverse evolution of the desired target state, providing a gradient towards improved performance.  The process continues until the forward and inverse trajectories match to an acceptable degree. 

Details of the optimal tracking algorithm applied to decoupling can be found in \cite{Neves09}, so they need  not be repeated here.  Briefly, the standard GRAPE algorithm \cite{GRAPE} is applied at each point where the time-domain signal is to be sampled rather than only at the end of the decoupling sequence.  Decoupling that is independent of the sampling rate \cite{BUSS} can be obtained by applying the tracking to each time-step of the RF waveform, which is typically much faster than the sampling rate.
The algorithm can't make $S_x=1$ at all times during the sequence, but asking it to do so forces it to search for the closest approach to this ideal. The result maximizes the average value of $S_x$ and minimizes deviations from the ideal value, which minimizes sidebands.  This very simple criterion, that $S_x=1$ as often as possible during the sequence, requires no input concerning the mechanism for achieving it. Yet the algorithm utilizes precisely the mechanism presented here, as it must.  Illustrations of the theory in practice are considered next.

\section{Representative Examples}
\label{sec:Examples}
To summarize so far, the simplest and most direct criterion for ideal decoupling is $S_x = 1$ (on-resonance S spins, neglecting relaxation) at all sample points during the acquisition.  Chemical shift evolution of the S spins can be ignored since it commutes with the other terms in the Hamiltonian and merely shifts the frequency of the decoupled signal.
Deviations from the ideal produce satellite lines (sidebands) whose intensity reflects the magnitude of the deviation.  RF applied to the I spins provides a mechanism for refocusing $S_x$ faster than the free coupling evolution period, $2/J$, by toggling the sign of antiphase magnetization $2S_yI_z$, which results in a time-dependent reduced coupling $\mathcal{J}_r(t)$.  The average reduced coupling $\bar{\mathcal{J}_r} = 0$ for any refocusing period.  

\subsection{Standard cyclic, periodic decoupling sequences}

High power inversion pulses provide a simple illustration of nearly ideal cyclic decoupling. The parameters for the net equivalent rotation at each time, derived from \Eq{U_pm}, are plotted in \Fig{DecParams} for 
on resonance irradiation of the I-spins by (i) two composite $90_x 180_y 90_x$ pulses and (ii) two adiabatic hyperbolic secant pulses.  
The terms $\beta_\pm = \omega_\pm t$, which can be interpreted as the net equivalent I-spin rotations at time $t$ for coupling offsets $\pm{\cal J}/2$, are approximately 360$^\circ$ at the end of the second pulse in each example, giving $\bm{\hat\omega}_+ \cdot \bm{\hat\omega}_- = 1$, so that $S_x = 1$ from the first term of \Eq{S_x(t)}.  Sampling only at the end of periodically applied cycles therefore gives a perfectly decoupled signal with zero contribution from the second term in the equation.  
\begin{figure*}[b]
\includegraphics[scale=.95]{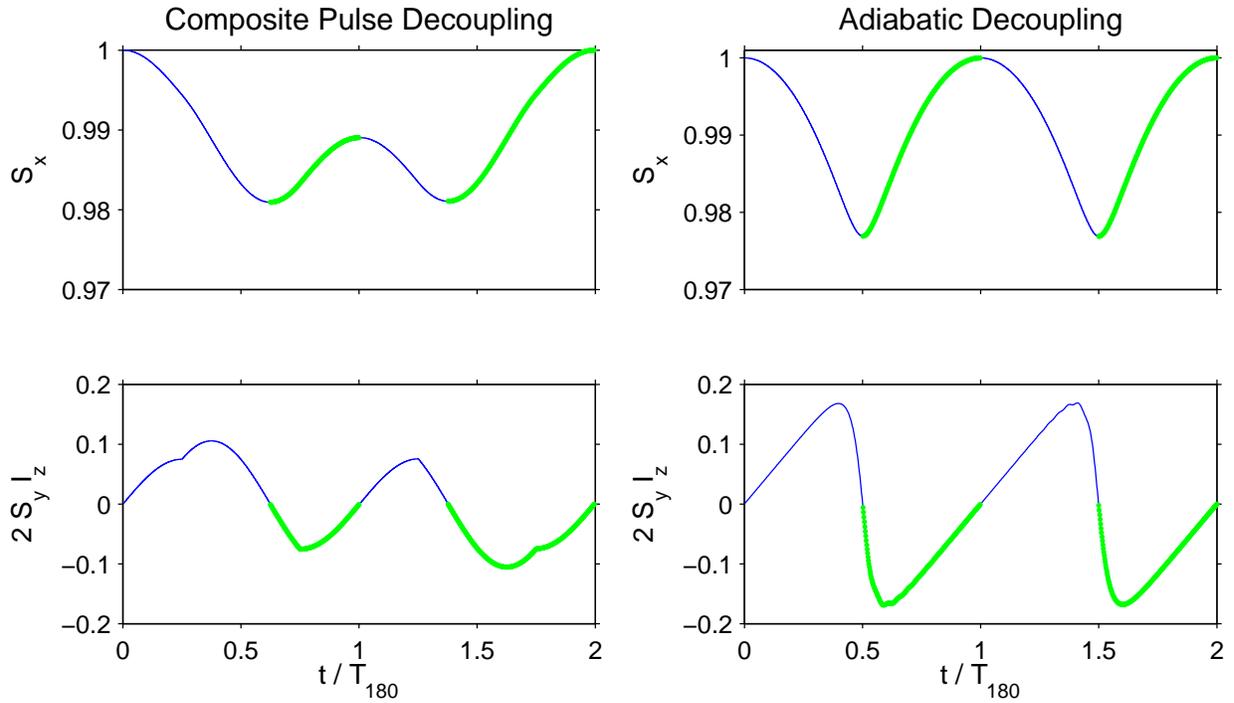}
\caption{The decoupling mechanism expressed through \Eq{SxdotEq} is illustrated for the pulses of \Fig{DecParams} applied to the I-spins of a coupled IS spin-1/2 system ($J=150$~Hz). The exact correlation between the $\mp$ slope of the calculated signal $S_x(t)$ (thin, blue/thick, green  lines, respectively) and the $\pm$ sign of $2S_yI_z$ is illustrated for \textbf{(left):} composite pulse decoupling and \textbf{(right):} adiabatic decoupling.  Cyclicity requires two inversion pulses to generate a net free I-spin rotation of 360$^\circ$ and refocus coupling evolution of $S_x$.  In addition, a single adiabatic inversion also refocuses $S_x$ by more efficiently generating the required antiphase magnetization.  In all cases, $\int 2S_yI_z = 0$ for a refocused period, as discussed in $\S$\ref{subsubsec:DecMech}.}
\label{CyclicDec}
\end{figure*}
The exact one-to-one correspondence between the sign of antiphase magnetization $2S_yI_z$ which determines the instantaneous coupling $\mathcal{J}_r(t)$ and the sense of the coupling evolution for $S_x$ is shown in \Fig{CyclicDec}.
The signal $S_x(t)$ decreases/increases in lock-step with positive/negative $2S_yI_z$.  At the end of the sequence, the integrated antiphase magnetization is zero, which is the condition that $S_x=1$.  
However, one only needs $\beta_+ \approx \beta_-$ and approximately equal rotation axes $\bm{\hat\omega}_\pm$ to obtain the same result, which occurs at the end of the first adiabatic inversion pulse, where $\int 2S_yI_z\,dt = 0$ as well, and $\beta_\pm \approx 180^\circ$.  
Thus, adiabatic sequences, designed only with cyclicity in mind, also utilize the possibilities for \textit{noncyclic} decoupling emphasized in \cite{Waugh82}.  
\begin{figure}[h]
\includegraphics[scale=.7]{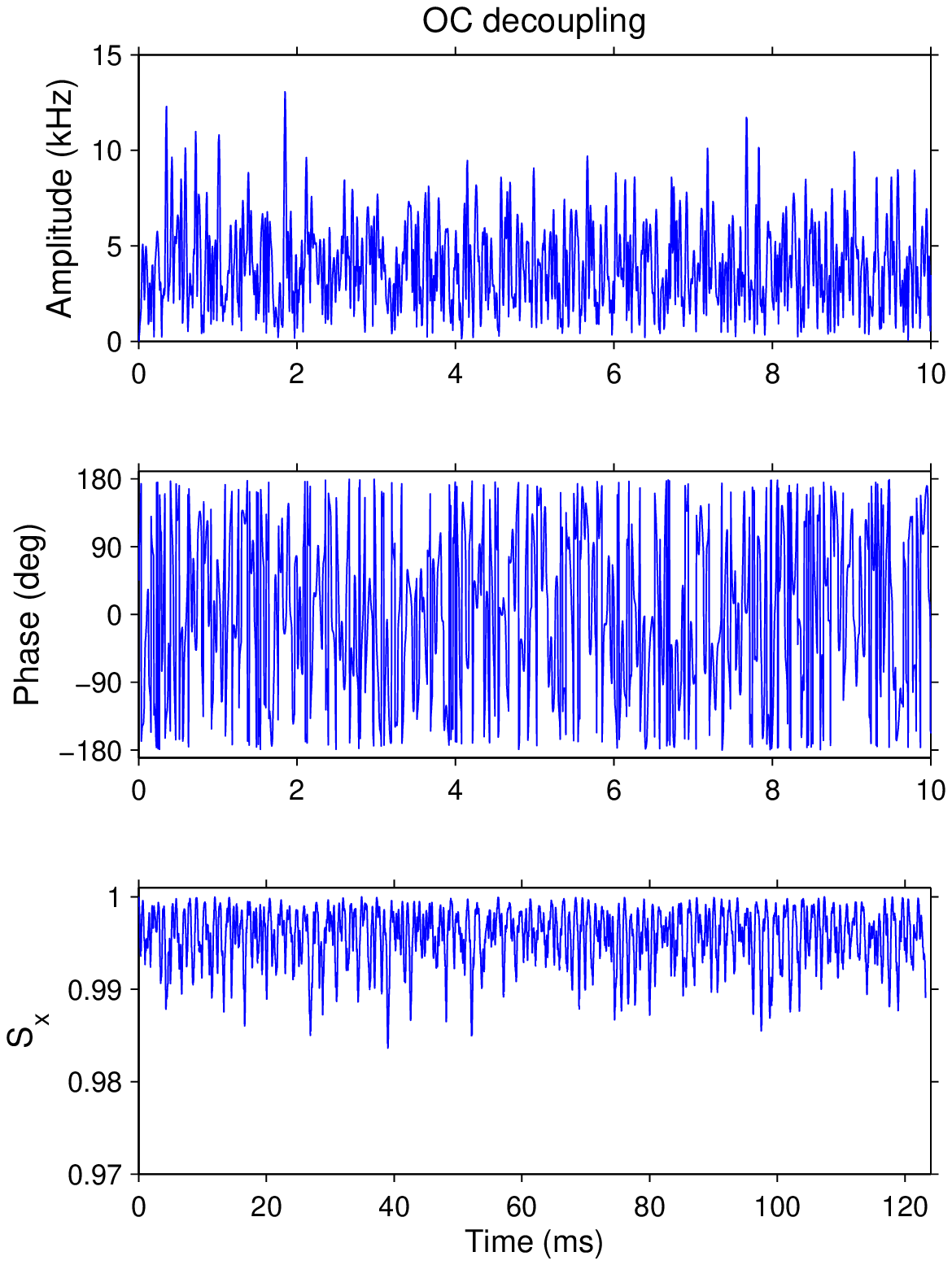}
\caption{RF amplitude and phase for a representative 10~ms segment of an aperiodic, non-cyclic 123.2~ms optimal control BUSS decoupling pulse (broadband uniform sideband suppression) \cite{BUSS}. The decoupled signal $S_x(t)$ is plotted for off resonance (10~kHz) irradiation applied to the I spins of an IS spin-1/2 system ($J=150$~Hz).  At a power level (rms $B_1$ = 4.4~kHz) satisfying the power constraints of modern cryoprobes, BUSS uniformly decouples a 45~kHz bandwidth with ultra-low sidebands of $\lesssim 0.2\%$ relative to the decoupled peak.}
\label{OCPulse}
\end{figure}
\subsubsection{Limitations of standard criteria}
\label{subsubsec:CyclicLimitations}
Adiabatic pulses illustrate one possibility for non-cyclic decoupling, refocusing S-spin magnetization with a single inversion pulse rather than the two inversions required for continuous-wave and composite pulse decoupling.
More generally, however, other than the cyclic condition $\beta_\pm \approx 0$ and $\bm{\hat\omega}_+ \approx \bm{\hat\omega}_-$ at the end of periodic sequences, there are  no simply defined criteria for designing propagators $U_\pm$ that refocus $S_x$ based on equivalent I spin rotations over a desired range of offsets.  Although the more general periodic constraint $\beta_+ \approx \beta_-$ is satisfied throughout the on-resonance sequences of \Fig{DecParams}, due to the small net offset $\pm\mathcal{J}/2$ relative to the applied RF, the corresponding rotation axes $\bm{\hat\omega}_\pm$ at each time are not necessarily equal when $S_x \approx 1$.  Near the beginning of the adiabatic sequence, $\bm{\hat\omega}_+ \cdot \bm{\hat\omega}_- \approx -1$ and 
$\beta_\pm = 1^\circ$, so the second term of \Eq{S_x(t)} is responsible for the result $S_x \approx 1$.  Near the end of the sequence at $t/T_{180} = 1.95$, 
$\bm{\hat\omega}_+ \cdot \bm{\hat\omega}_- \approx 0$.  The axes are orthogonal, and $\beta_\pm \approx -2.2^\circ$, so 
$\cos\frac{1}{2}(\beta_+ \pm \,\beta_-)\approx 1$ and both terms contribute almost equally to the signal $S_x \approx 1$.  

In addition, there are many other combinations of parameters that result in $S_x \approx 1$ during the sequence, with a multitude of additional possibilities for off-resonance decoupling.  These simple examples illustrate the difficulty in defining I-spin criteria for broadband decoupling of the S-spins.  The default has thus been the simple cyclic criterion.  However, the analysis also indicates there is considerable latitude for designing \textit{nonperiodic} sequences.  There are many more opportunities for achieving a decoupled $S_x \approx 1$ signal when unconstrained by traditional cyclic/periodic approaches.
\begin{figure}[h]
\includegraphics[scale=.7]{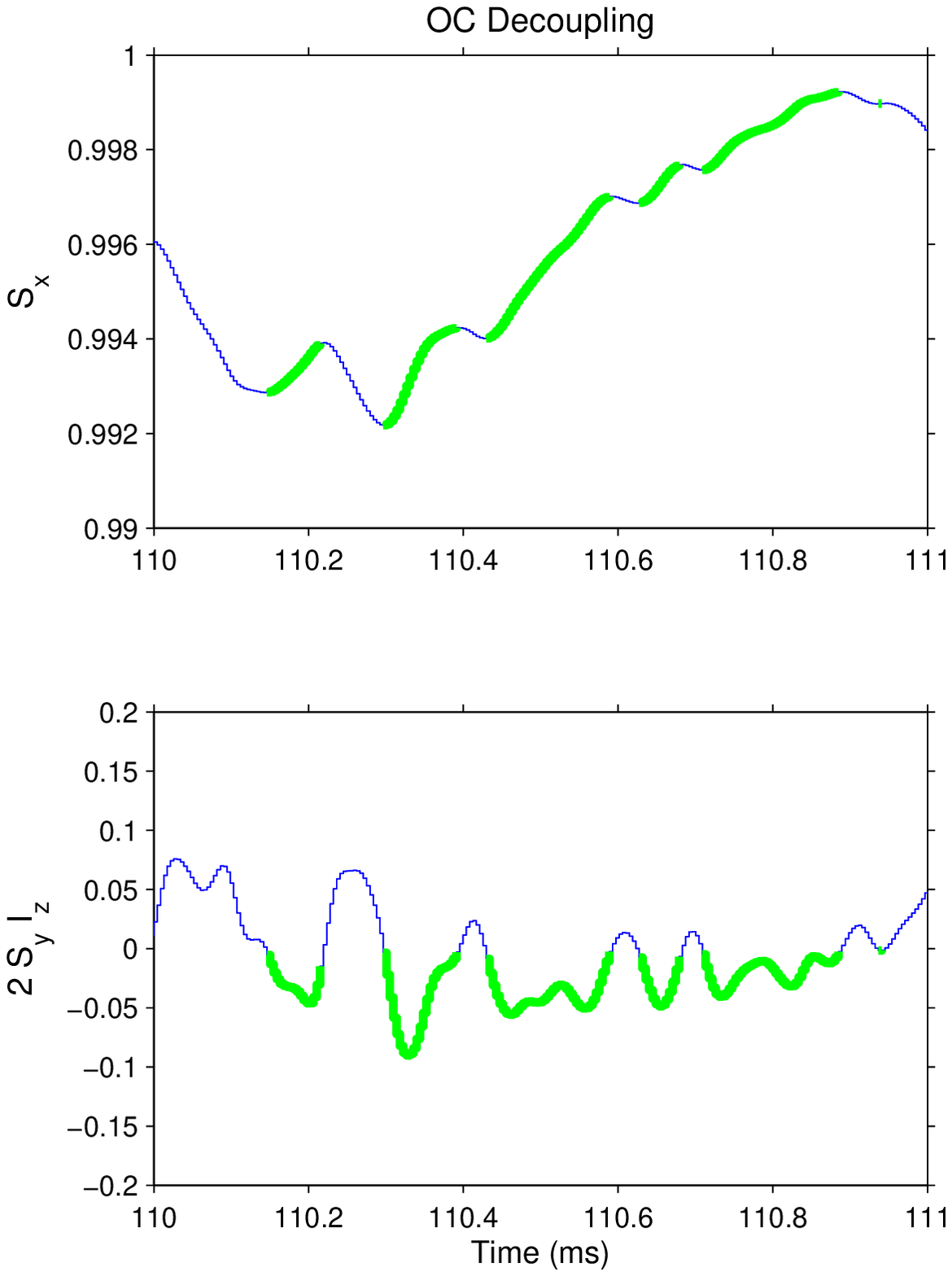}
\caption{The decoupling mechanism expressed through \Eq{SxdotEq} is illustrated for a representative 1~ms segment of the aperiodic, non-cyclic optimal control BUSS pulse of \Fig{OCPulse}.  Signal $S_x$ decreases for $2S_yI_z > 0$ (blue, thin line), with refocused precession and increasing $S_x$ for $2S_yI_z < 0$ (thick, green line).}
\label{OCdec}
\end{figure}
\subsection{Aperiodic decoupling}

The fundamental criterion for good decoupling is the invariance of S spin magnetization over the desired bandwidth.  This simple and direct criterion can be readily enforced by the aforementioned optimal control algorithm via the decoupling mechanism presented here.  Yet, the mechanism itself does not have to be either known or included in the algorithm. 
A target state $S_x=1$ at each time throughout the acquisition period drives the iterative optimal tracking procedure towards solutions which reverse the sign of $2S_yI_z$ as often as possible within the constraints imposed by the given bandwidth and peak RF. The concurrent production of two-spin coherence is the mechanism by which the coupling is reduced.

\Figure{OCPulse} shows the amplitude and phase of a low power 
(rms $B_1$ = 4.4~kHz) broadband uniform sideband suppression (BUSS) decoupling pulse designed using optimal control \cite{BUSS}.  This 123.2~ms pulse uniformly decouples a bandwidth of 45~kHz with ultra-low sidebands of 
$\lesssim 0.2\%$ relative to the decoupled peak.  The aperiodic and acyclic evolution of the signal, with $S_x \approx 1$ throughout the sequence, is shown in the bottom panel of the figure.  The exact correlation between the $\mp$ slope of $S_x(t)$ and the $\pm$ sign of $2SyIz$ is shown for a representative 1~ms segment of the decoupling pulse in \Fig{OCdec}.  

Although optimal control is also capable of deriving inversion pulses and phase cycles that improve decoupling within the cyclic framework, ultimate performance results from a more global approach that optimizes the entire sequence as a single unit, as in \cite{Neves09, BUSS}.

\section{Conclusion}
RF irradiation of the I spins in a heteronuclear spin-1/2 IS system decouples the S spins by reversing the sign of antiphase magnetization, $2S_yI_z$, to refocus the S-spin coupling evolution.  An exact expression for the resulting reduced coupling, \Eq{J_r}, shows the crucial role also played by the timely creation of two-spin coherence, $2S_yI_{x,y}$, in efficient decoupling.  In its absence, $S_x$ evolves at the natural coupling frequency.  Both 
$\int 2S_yI_z\,dt$ and the average reduced coupling, \Eq{AvgJ_r}, are zero over any period in which $S_x$ is completely refocused.  The best decoupling and minimal sidebands result when the longest refocusing period (including partial or incomplete refocusing) is the shortest possible. 

Formerly, S-spin decoupling was approached from the perspective of free I-spin rotations, requiring cyclic and/or periodic sequences that were unnecessarily restrictive and limited.  The defining criterion for ideal decoupling, that the magnetization $S_x = 1$ at all sample points during the acquisition, was not directly addressed.  The multitude of additional possibilities for achieving $S_x = 1$ were, thus, underutilized.  Optimal control provides an efficient means for, in effect, inverting the theory and accessing the full potential of the decoupling mechanism to produce significantly improved decoupling sequences, freed from former constraints of cyclic and periodic sequences. An exact, general theory for understanding the mechanics of decoupling and a procedure for deriving optimal decoupling sequences provide a reasonably complete solution to the decoupling problem.

\begin{acknowledgments}
The author gratefully acknowledges support from the National Science Foundation under Grant CHE-1214006.
\end{acknowledgments}

\bibliography{bib-database-TES}

\end{document}